\documentclass[conference,compsoc]{IEEEtran}

\usepackage[nocompress]{cite}
\usepackage{url}
\usepackage{array}
\usepackage[caption=false,font=footnotesize,labelfont=sf,textfont=sf]{subfig}
\usepackage{fixltx2e}
\usepackage[pdftex]{graphicx}
\usepackage{slashbox}
\graphicspath{{img/}}
\DeclareGraphicsExtensions{.pdf,.jpg,.png}

\begin{document}

\title{Autonomous smartphone apps:\\self-compilation, mutation, and viral spreading}

\author{
  \IEEEauthorblockN{Paul Brussee}
  \IEEEauthorblockA{Delft University of Technology\\Email: P.W.G.Brussee@student.tudelft.nl}
  \and
  \IEEEauthorblockN{Johan Pouwelse}
  \IEEEauthorblockA{Delft University of Technology\\Email: J.A.Pouwelse@tudelft.nl}
}

\maketitle

\begin{abstract}

We present the first smart phone tool that is capable of self-compilation, mutation and viral spreading.
Our autonomous app does not require a host computer to alter its functionality, change its appearance and lacks the normal necessity of a central app store to spread among hosts.
We pioneered survival skills for mobile software in order to overcome disrupted Internet access due to natural disasters and human made interference, like Internet kill switches or censored networks.
Internet kill switches have proven to be an effective tool to eradicate open Internet access and all forms of digital communication within an hour on a country-wide basis.
We present the first operational tool that is capable of surviving such digital eradication.

\end{abstract}

\begin{IEEEkeywords}
self-compilation, mutation and viral spreading.
\end{IEEEkeywords}

\section{Introduction}

Natural disasters and human made disruptions can both be the cause of communication systems failure.
Whenever a natural disaster occurs communication infrastructure is often completely destroyed or affected in such a way that it becomes unusable for longer periods of time.
Terrible loss of life has been blamed directly on the breakdown of cell phone service in the stricken area after the earthquake in Haiti in 2011 \cite{denis2010haiti}.
The infrastructure in the capital had collapsed with the buildings and remaining networks were out of range and quickly overloaded.
The landfall of hurricane Katrina in 2005 caused a communication blackout and large scale network outages lasting for multiple weeks \cite{renesys2005katrina}.
Besides the flooding, the continuing power outages disrupted Internet access for months.
An unstable electrical grid has caused multiple large scale blackouts in the United States in recent history \cite{cnn2003poweroutage}.
Without a stable electricity supply many forms of modern communication are useless, because they depend on infrastructure that has to be powered continuously to function without sufficient backup power.
Not just highly developed countries have to deal with a troubled energy supply.
Especially the weak grid in Africa hurts the economy of countries like Ghana, Malawi and Tanzania \cite{nyt2015afrika}.

Besides the natural causes and indirect human causes just mentioned, there are also multiple known cases of various intentional communication blackouts orchestrated on a state level.
For instance Cuba is barely connected to the Internet all, prompting its citizens to come up with offline Internet \cite{watts2014havana}.
The Great Firewall of China is the well known name of all the layers of censorship that the Chinese government enforces in all their territories \cite{hrw2006china}.
This effectively causes a communication blackout that affects all outbound connections from within the borders of China.
The government of Iran blocks almost 50\% of the world’s top 500 most visited websites \cite{halderman2013iran}.
In Turkey social media were completely blocked for almost a day due to a temporary broadcast ban following a deadly attack in the capital \cite{passary2015socialmedia}.
These are just a few examples of how a single actor can cause communication systems to fail to function.

\subsection*{Problem Description}

We aim to address the problem of robust and resilient mobile systems, specifically usable without exiting infrastructure and consisting of innocuous hardware \& software; moving toward a censorship-free Internet: \cite{pouwelse2012censorshipfree}
\begin{itemize}
\item{The adversary can observe, block, delay, replay, and modify traffic on all underlying transport. Thus, the physical layer is insecure.}
\item{The adversary has a limited ability to compromise smart phones or other participating devices. If a device is compromised, the adversary can access any information held in the device\'s volatile memory or persistent storage.}
\item{The adversary cannot break standard cryptographic primitives, such as block ciphers and message-authentication codes.}
\end{itemize}

\section{Related work}

Various projects attempted to provide at least a partial solution to the problem just described.
Beginning with the fact that any network which has a single point of failure is vulnerable.
For example a mobile network that covers an area with a single cell tower, or wherever in dense cities there are tall buildings blocking the signals of all but one tower.
Even redundant infrastructure becomes useless when the switch point is wiped out or otherwise disabled.
That is one of the reasons why Internet switches are critical points for wide spread outages.
Additionally such central points also have been used in Egypt in 2011 \cite{renesys2011egypt} and Syria in 2012 \cite{renesys2012syria} to quickly terminate all Internet access in a country intentionally.
Encrypted tunneling and anonymous proxy technology is available \cite{psiphon2015} to circumvent some forms of blockades but uses existing infrastructure, thus does not solve the problems just mentioned.
The challenge is now to use the pervasive smart phone phenomenon to create a resilient alternative Internet infrastructure.

Mesh networks are a solution to this problem because every wireless device effectively becomes a cell tower and switch itself, thus providing connectivity service wherever it goes, as long as it stays within reach of any other mobile device connected to the mesh network; otherwise it will be fairly limited connectivity.
Now there is no single point of failure anymore as all devices have this cell tower capability and could therefore take over for each other.
This way the network is protected against damage to infrastructure, because that is not required anymore, as well as power outages or an unstable electricity supply, because wireless devices typically carry batteries themselves.
For example the power bank phone carrying a 10000mAh battery capable of charging other devices as well via USB \cite{quartey2015phone} is still small enough to be carried on person, so it will survive a natural disaster together with its user.
Compare this to a current smart phone having a capacity of less than a third than that, around 3000mAh.
The phone is popular in Ghana, which is not surprising since the capital regularly suffers from blackouts lasting more than 36 hours.
Mesh networks can differ significantly in the way they are set up.
For instance they may grow organically without oversight or they may be deployed and extended in an organized manner by a central authority.
Meraki is an example of the latter kind \cite{cisco2015meraki}.
Having a central authority re-introduces the problem of human made failure of the network.
To overcome this problem various communities have formed around projects that are deploying DIY-mesh-networks in multiple cities around the world.
For example Guifi.net in Spain \cite{guifi2011mesh}, the Free Network Foundation \cite{tfnf2011mesh} and Commotion \cite{commotion2012mesh} in America, FreiFunk \cite{freifunk2014mesh} in Germany and the Serval project \cite{serval2015mesh} in New Zealand.
All of these projects offer software to run on one\'s own router or smart phone to join the mesh network, subsequently providing other wireless devices access via said router or phone, effectively extending the mesh network.
In addition to that the Freedom Network Foundation also offers a hardware design for the FreedomTower to improve performance of the network in the entire neighborhood, and even a lab in two separate data centers open for anyone to find out and evaluate the best configuration of hardware and software for connectivity on a larger scale.
The general aim of these projects is to provide a robust and resilient censorship free communication network anywhere the users are, unaffected by the lack of communication infrastructure. 

Having the same general aim here are a few specific applications for smart phones using Bluetooth and WiFi to implement the mesh network:
\begin{itemize}
\item{Open Garden: Internet Sharing \cite{opengarden2015mesh}.}
\item{FireChat \cite{opengarden2015firechat} also from Open Garden; chat only.}
\item{Bleep \cite{bittorrent2014bleep} from BitTorrent; supports chat and voice-calls.}
\item{ShadowTalk \cite{peersafe2015shadowchat} from PeerSafe; supports chat and voice-chat with time-lock.}
\item{Serval Mesh \cite{serval2015mesh} from the Serval Project; supports file transfers in addition to chat and voice-calls.}
\item{Tuse \cite{tuse2015} from Tuse; supports sharing the app p2p in addition to file transfers, chat and voice-calls.}
\item{Twimight \cite{twimight2012mesh} for using Twitter offline until it finds an uplink to Internet.}
\end{itemize}
All but the last two of these apps support public messaging and private messaging with end to end encryption.
All of them are free to use and available on the Google app store.
However that central app store is, by definition of it being central, vulnerable to the same problems as mentioned before.
Either a decentralized app store is required, like tsukiji \cite{michael2015tsukiji}, or the possibility of downloading and installing the app without the use of an app store all together.
This process is called side loading and is possible on the Android platform.
However few apps are capable of virally spreading.
Apps that could benefit greatly from using such technology are the eyeWitness app \cite{bbc2015eyewitness}, from the Eye Witness project, because a tactic to safeguard a user from an oppressive entity would be to make sure multiple copies of eyewitness evidence are virally spread as fast and easy as possible.

Although multiple IDE apps are available for Android \cite{idr2014ide}, none of them can replicate themselves like a Quine and neither are they capable of autonomously mutating their source.
An earlier app from Delft University of Technology called DroidStealth \cite{tudelft2014droidstealth} has very limited mutation capability, only the binary form and not the source code can be mutated, meaning that the functionality cannot be changed, only the app name and icon, however it can virally spread via Android Beam.

A polymorphic computer virus avoids detection by mutating itself \cite{nachenberg1997antivirus}.
Each time a new host is infected the binary software is altered to look completely different, yet remains unchanged in functionality.
Polymorphic viruses are an established technique, dating back before 1997.
These mutating viruses do not have access to their own source code and lack an embedded compiler, making self-compilation more generic and likely extra powerful.

\section{Autonomous App Design}

The following design aspects are aimed to accomplish the criteria set out in the problem description: a robust and resilient mobile system.
Smart phones have multiple wireless technologies on board, like Bluetooth and WiFi, carry a battery and are very widespread all around the world, making it the perfect platform to use for a mesh network.
Therefore to be most accessible the software will need to run on the open source platform with the largest install base: Android.

Mobile service providers often modify the operating system and do not grant users administrative privileges on their own device for whatever reason.
Gaining root access to take these administrative privileges by force is possible, but not without the help of a development computer.
This means that the app must not require root access to the phone in order to be autonomous.

\subsection{Viral Spreading}

\begin{figure}[h!]
\centering
\includegraphics[trim=1cm 6cm 6cm 1cm, clip=true, width=3.5in]{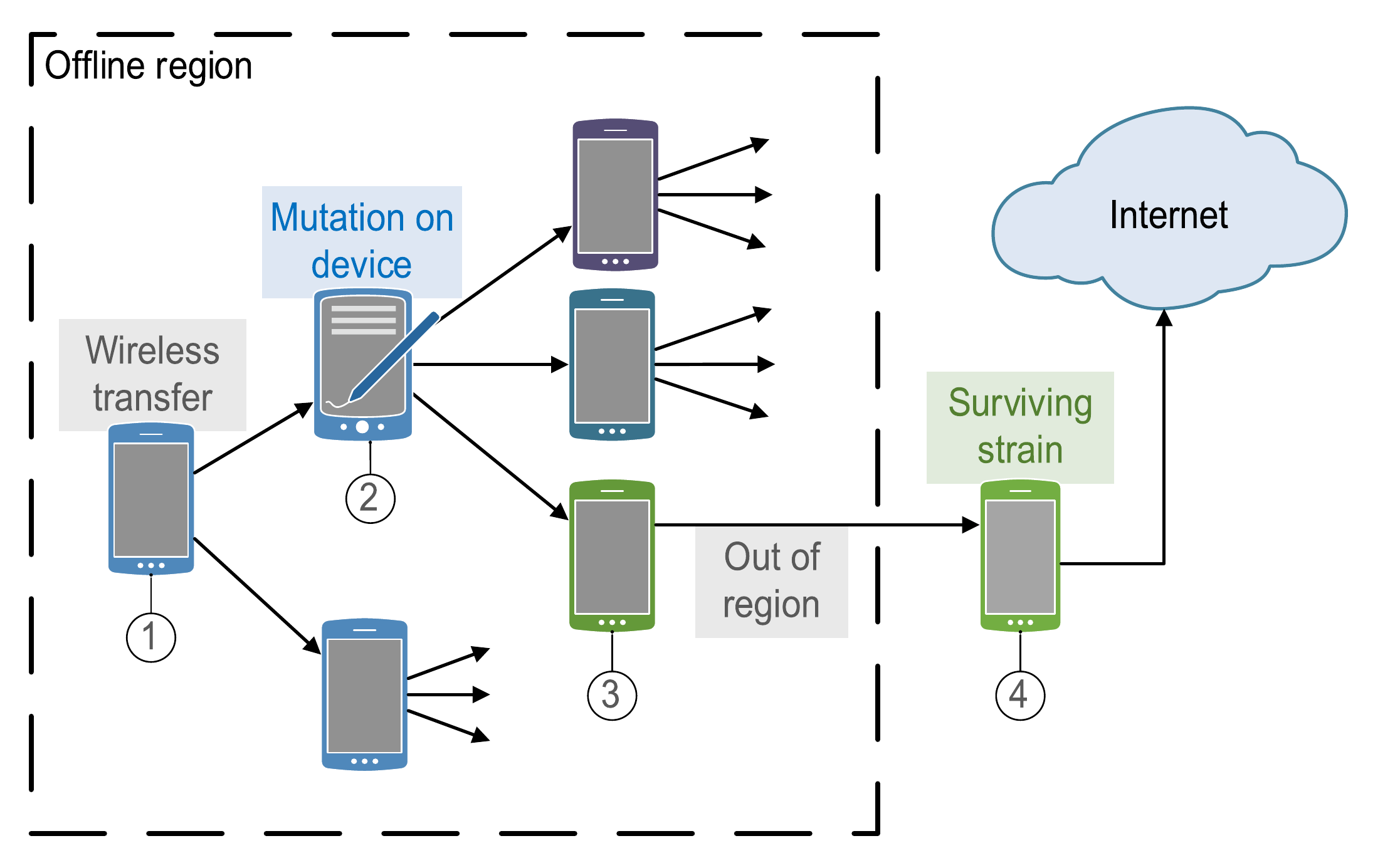}
\caption{ 1) P2P wireless transfer via NFC and Bluetooth to any number of devices. 2) Any device can mutate the app and distribute without limitation. 3) One strain of the app travels out of the offline region. 4) The surviving strain is connected to the rest of the world.}
\label{fig_regions}
\end{figure}

Making the app capable of outputting itself and sending it to another device eliminates the normal necessity of a central app store and makes it capable of spreading like a virus.
This bio-inspired mutation and spreading requires modifying the source code and the capability of running on a variety of host hardware, as described in the next subsection.
Figure \ref{fig_regions} shows viral spreading from within an offline region until it reaches an area with unrestricted Internet access.
Building upon existing trust networks, like family or business partners, it becomes evident that interconnection of local clusters is key.

\subsection{Self-compile}

While virally spreading the app may encounter different platforms, like a new Android version or hardware architecture, so the app needs to be able to re-compile for such a new environment.
In addition the app needs to solve dependencies for new environments that may be unavailable at the time the app needs to re-compile so it should contain all build tools and libraries that may be required during the entire Android build process.

Usually Android apps are developed on a computer with a software development kit (SDK) and integrated development environment (IDE).
To make the app fully function without the need for such a system requires integrating the SDK and entire build chain into the app.
Effectively it requires an embedded app factory consisting of all the required build tools, chained together.
Various levels of self-compilation can be discerned: 1. the source code 2. the 3rd party libraries 3. the build tools.
Compiling the non-Java tools and C/C++ libraries would require the native development kit (NDK) in addition to the SDK.

Because it is possible for the user to user to add and remove any content in the installation package the app is capable of serving as an alternative communication network.

\subsection{Mutate}

To increase the survival skills of the app in adverse conditions it must be able to be modified immediately when needed on the device itself.
Crowdsourcing would be a transparent and decentralized way of improving the survival skills of the app.
Being able to integrate other pieces of code, resources and new or modified libraries into the app makes it incredible versatile and adjustable to any situation.
Periodic mutation would prevent a single signature of the app to be made in order to block it.

\begin{figure}[h!]
\centering
\includegraphics[width=3in]{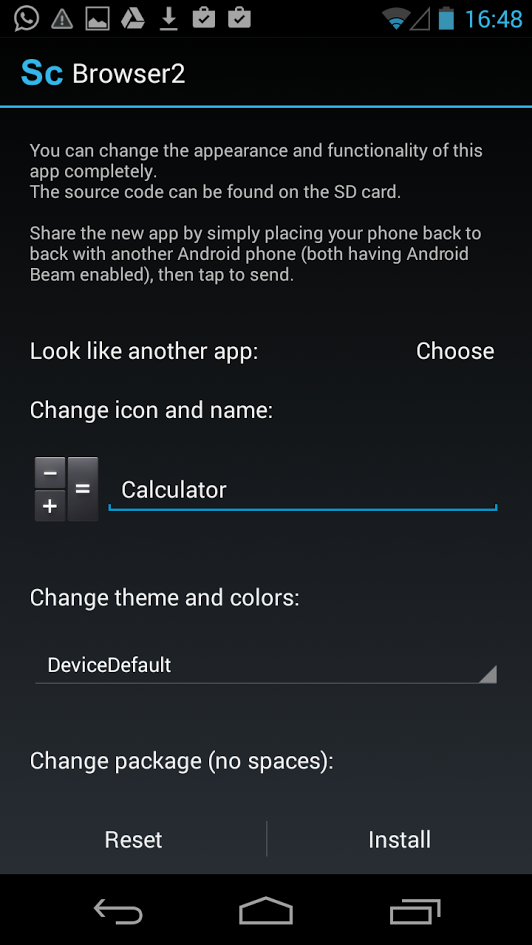}
\caption{User interface for mutation.}
\label{fig_screenshot}
\end{figure}

The user has full access to and control over the source code, but needs to be aware of possible compiler injection by the running instance of the code, especially in case the app was received from an untrusted source.

Leveraging the power of reflection in code the app could indicate and perhaps look for new components itself that are available in the mesh network and become even more resilient through continuous self-modeling.
The mutation is aimed at creating generations of the app with increased fitness.

\subsection{Innocuous Components}

Using innocuous hardware and software is especially important in case of a dissent network \cite{hasan2013dissent}.
A casual search or monitoring may not pick up on an app that looks like a calculator or is generally inconspicuous.
Also separate pieces of code may be innocuous on their own, so it is only a matter of putting these together.
A game could for example be embedded with a special launch pattern to open the encrypted content within.

\subsubsection*{Anonymity vs. Trust}

Removing usage tracks and signing the application with the default debug key makes it hard to trace back to an individual user.
In case of illegal embedded content this could provide author, publisher and reader anonymity, but not server, document and query anonymity \cite{freehaven2009anonymous}.
The trustworthiness however of any true anonymous source is unverifiable, so transitive trust is desirable.

\subsubsection*{Plausible Innocence}

To be able to use IDE features if the user is not a software developer may be tricky.
However this depends on the many different use cases for non-dissent purposes of the app, so the usage is not incriminating in and of itself.
As technology only amplifies human intent \cite{hasan2013dissent} having IDE features on your phone does not create political movement.
Like a hammer it can be used for good or evil and will find its role in the social network of people that is already established.
To minimize use for harm various attack vectors, like staining, need to be considered.

\section{Implementation}

We fully implemented our autonomous app design.
The source code of the SelfCompileApp is available on GitHub \cite{brussee2015selfcompile}.

\begin{figure}[h!]
\centering
\includegraphics[trim=4cm 0.5cm 4cm 0.5cm, clip=true, width=3.5in]{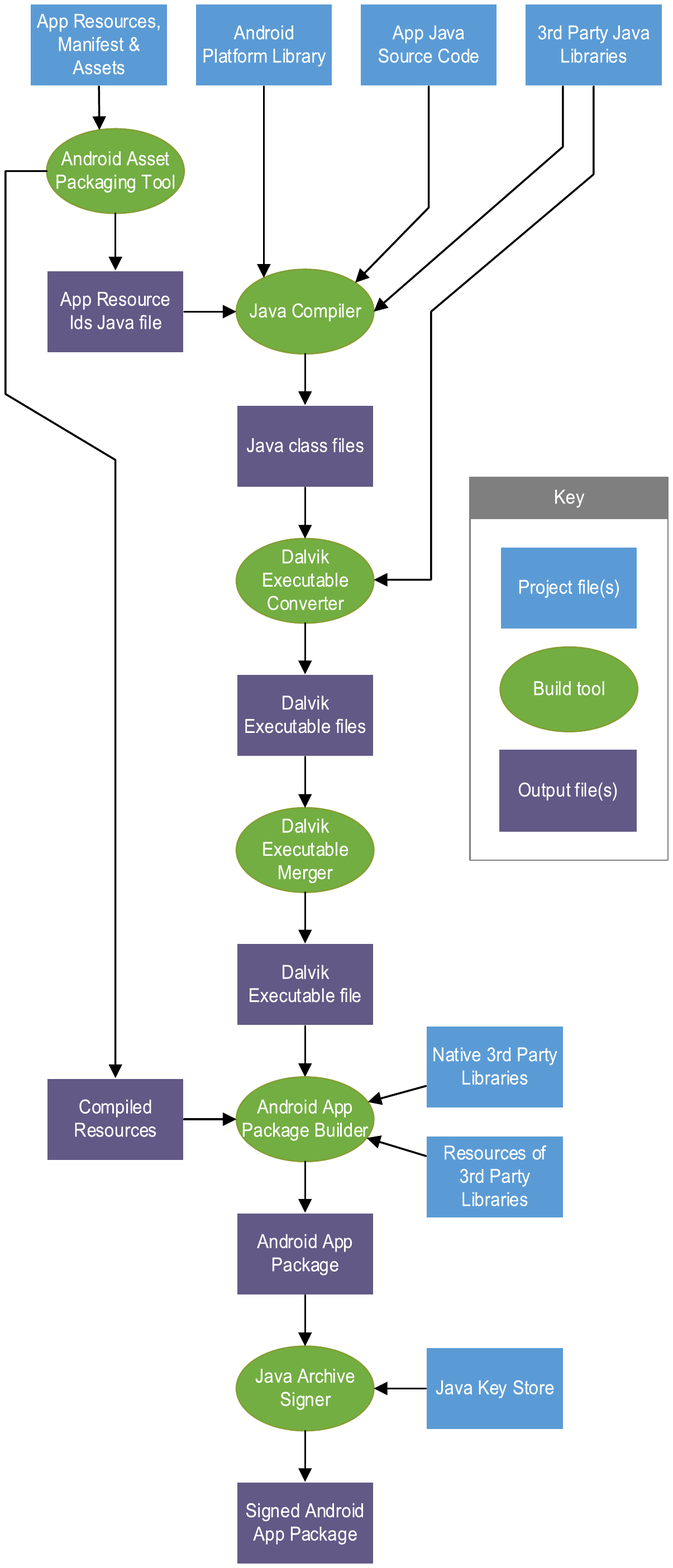}
\caption{Implemented build chain.}
\label{fig_small_chain}
\end{figure}

As described in the previous chapter the Android SDK and platform library are required for building an installable app package (.apk).
Figure \ref{fig_small_chain} shows the implemented tool chain that is required to let the app self-compile on the device.
The Android Open Source Project (AOSP) made most of the build tools source code available online \cite{google2015source}.
All tools except the Android Asset Packaging Tool (aapt) are implemented in Java, which is convenient since Android apps can be written in Java as well.

\subsection{Android Asset Packaging Tool}

The Android Asset Packaging Tool compiles the app manifest and resources like images, layout and language .xml files into one binary file and generates R.java.
This file contains all the IDs of the compiled resources that can be referenced from the app source code.
Aapt then packages the binary resources and all files in the assets folder into a preliminary .apk file.

Since aapt is not implemented in Java it requires the NDK to build it for any CPU instruction set(s) you want to support, which is part of the extended Android build process \cite{ducrohet2015buildworkflow}.
The app JavaIDEdroid showed this works for Android 4.4 and older \cite{arn2011javaidedroid} and proved to be compatible with Android 5.0 by dropping the resources for ultra high density screens because of the used older version of aapt.
All required native shared libraries are bundled into one native shared library file (.so) and included in the app installer package.
Because of a missing dependency CynogenMod was used for all tested devices running an Android API level higher than 18.

To remove the NDK dependency completely only aapt would need to be implemented in Java.
This allows for easy viral spreading to new hardware platforms and make the app fully capable of mutating all the dependencies and build tools, without the need for the GNU Compiler Collection (GCC) which TerminalIDE uses \cite{terminalide2013}.

\subsection{Java Compiler}

The Eclipse Compiler for Java (ECJ) is chosen because it is implemented in Java, so it can be easily included as a library in the app but also makes the compiler capable of compiling itself.
Currently all build tools are just included in the app as a library, but could be build from source just like the app source code.

\begin{figure}[h!]
\centering
\includegraphics[width=3in]{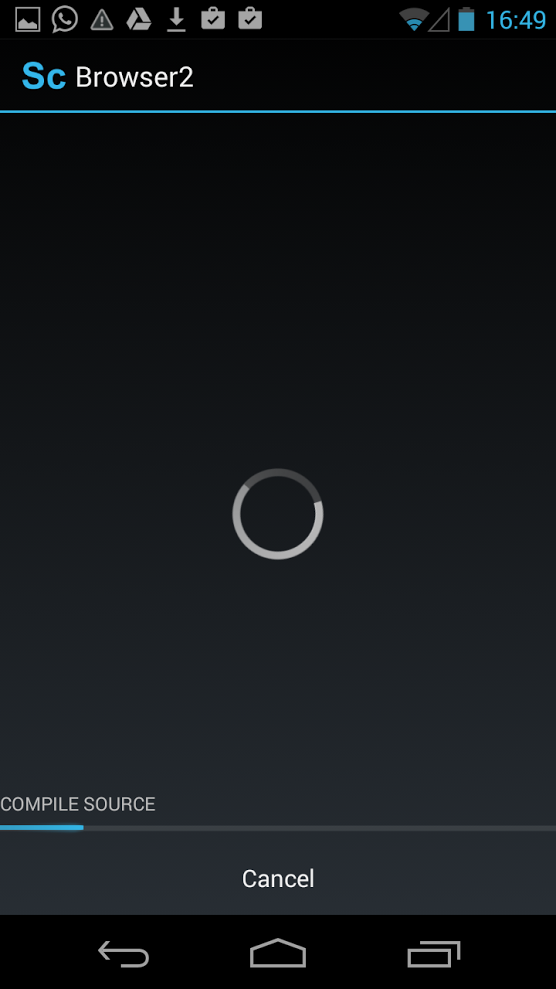}
\caption{Build in progress.}
\label{fig_screenshot}
\end{figure}

\subsection{Dalvik Executable}

After compiling the source to Java bytecode it needs to be converted into Dalvik bytecode because Android used to run Java in a process virtual machine (VM) called Dalvik.
In Android 5.0 this VM was replaced by an application runtime environment called Android Runtime (ART) which is backwards compatible and still uses the exact same Dalvik executable files (.dex).
This conversion process with the dx tool proved to require a huge amount of RAM, or the tool contains a serious memory leak.
Fortunately it is possible to convert the libraries separately so they are cached and only converted again if the hash of the .jar file has changed.
The source code is converted into one .dex file as well and subsequently merged with with the converted libraries.
Both the dx tool and DexMerger tool are implemented in Java and available from AOSP \cite{google2015source}.

\subsection{Android App Package Builder}

All compiled resources, native libraries, compiled files and asset files are combined by the ApkBuilder tool that also implemented in Java and available from AOSP \cite{google2015source}.
This includes the Android.jar platform library, required for compiling the Java source code, and the Java KeyStore, required for the last link of the build chain.
The ApkBuilder has the capability of signing the .apk itself, but does not support Java KeyStore, so a user would not be able to use or create a private key store without additional tools.
The uncompiled resources, manifest, source code, libraries and pre-dexed libraries are zipped and also added to the .apk.

\subsection{Package Signer}

As long as the same package name is used and the .apk is signed with the same certificate as a previous installation of the app it can be seamlessly updated.
Otherwise the mutated app will be installed alongside the other version, or fail to install if the package name is the same but the certificates do not match.
The tool ZipSigner \cite{ellinwood2013zipsigner} is used to sign the .apk file that can use any Java KeyStore, one of which containing the default Android debug key is included in the app.
As long as side loading is enabled on the Android device, which is necessary to be able to install the .apk file at all, the use of a debug key does not pose any problems.

\subsection{Wireless Transfer}

To spread the app trough the air the wireless technology of Android Beam is used that leverages the ease of use of near field communication (NFC) for easy connection set up of a Bluetooth large file transfer.
The user only needs to hold the backs of two devices together and tap the screen while the app is running to transfer it.
Bluetooth is then enabled on both devices, the devices are paired and Bluetooth gets disabled again after the transfer is complete, all completely automatic.

\section{Performance Analysis}

\begin{figure}[h!]
\centering
\includegraphics[trim=1.5cm 6.5cm 1cm 6.5cm, clip=true, width=3.5in]{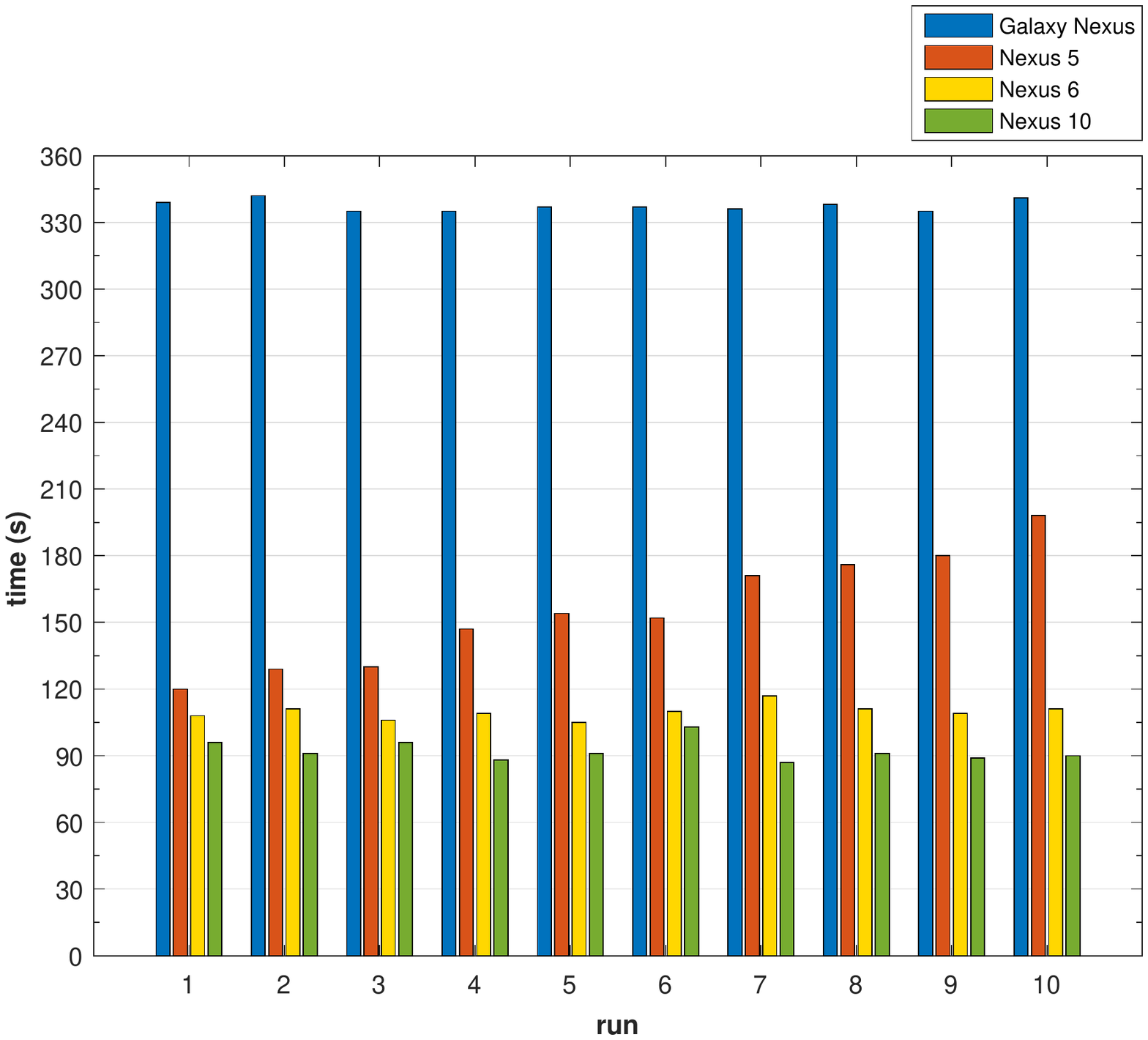}
\caption{Time to self-compile the app into an installable package.}
\label{fig_build_speed}
\end{figure}

We measured both the time to self-compile the app into an installable package (.apk) and also measured the time to wireless transfer the .apk to different devices like in a typical viral spreading scenario.
Figure \ref{fig_build_speed} shows the build time for 10 consecutive runs on four different devices.

Very notable is the significant difference in speed between the Galaxy Nexus and the other devices.
Besides the different hardware capabilities, this can be partially explained by the difference in how the Dalvik bytecode is executed.
As described in the previous section, Android used to run Java in a process virtual machine (VM) called Dalvik that performed just-in-time (JIT) compilation.
In Android 5.0 this VM was replaced by an application runtime environment called Android Runtime (ART) that performs ahead-of-time (AOT) compilation.

Also very notable was the apparent thermal throttling of the Nexus 5 device.
After a period of letting it cool down the results returned to the level of the first couple measurements.

The wireless transfer speed of four different Android Beam and Bluetooth large file transfer enabled devices was tested by sending over the .apk of 30.1MB.
For each pair of the four devices the averages of 3 consecutive transfers are shown in table \ref{tab_transfer_time}.
Two cells are left blank because only one Nexus 5 and one Nexus 10 device were at our disposal.
Notable is the slow transfer speed of only the Nexus 10.

\begin{table}[h!]
\renewcommand{\arraystretch}{1.5}
\centering
\begin{tabular}{l|*{4}{c}}
\backslashbox{From:}{To:}	&	Galaxy Nexus	&	Nexus 5	&	Nexus 6	&	Nexus 10	\\
\hline
Galaxy Nexus	&	227			&	221		&	209		&	419	\\
Nexus 5		&	211			&			&	149		&	360	\\
Nexus 6		&	198			&	147		&	139		&	357	\\
Nexus 10		&	409			&	400		&	359		&	\\
\end{tabular}
\vspace{10pt}
\caption{Average transfer time in seconds of 3 consecutive transfers per device pair.}
\label{tab_transfer_time}
\end{table}

\section{Conclusion \& Future Work}

Our work shows that it is possible for an Android app to be autonomous in recompiling, mutating and virally spreading over multiple hosts all without the need for a host computer, root permissions and an app store.
Therefore our work provides a robust and resilient network and platform to overcome disruptions due to natural disasters or made made interference.
Since the user can add and remove any content in the app a social media experience does no longer have to be hampered by Internet kill switches as it is now.

The app currently can spread among Android devices and as a next step could be considered cross-compiling for the Android, iOS and Windows Phone platforms.
Continuous self-modeling through code reflection and higher levels of self-compilation, by recompiling the 3rd party libraries and build tools themselves, are left for future work.

A point for consideration is the minimization of the use for harm of the app, and the risk for harm by use of the app.
Various attack vectors could be considered to improve the fitness of the app and lower the risk of usage in a dissent network use case.

\IEEEtriggeratref{24}

\bibliographystyle{IEEEtran}
\bibliography{IEEEabrv,SelfCompileApp}


\end{document}